\documentclass[twocolumn,letterpaper,aps,prl,
superscriptaddress,showpacs,amsmath]{revtex4}
\usepackage{graphicx}
\newcommand{\vc}[1]{\boldsymbol{#1}}

\begin{document}
\title{On the Origin of Zigzag Magnetic Order in Iridium Oxide Na$_2$IrO$_3$}

\author{Ji\v{r}\'{\i} Chaloupka}
\affiliation{Max Planck Institute for Solid State Research,
Heisenbergstrasse 1, D-70569 Stuttgart, Germany}
\affiliation{Central European Institute of Technology,
Masaryk University, Kotl\'a\v{r}sk\'a 2, 61137 Brno, Czech Republic}

\author{George Jackeli}
\altaffiliation[]{Also at Andronikashvili Institute of Physics, 0177
Tbilisi, Georgia}
\affiliation{Max Planck Institute for Solid State Research,
Heisenbergstrasse 1, D-70569 Stuttgart, Germany}

\author{Giniyat Khaliullin}
\affiliation{Max Planck Institute for Solid State Research,
Heisenbergstrasse 1, D-70569 Stuttgart, Germany}

\begin{abstract}
We explore the phase diagram of spin-orbit Mott insulators on a honeycomb 
lattice, within the Kitaev-Heisenberg model extended to its full 
parameter space. Zigzag-type magnetic order is found to occupy a large part of
the phase diagram of the model, and its physical origin is explained as due to 
interorbital $t_{2g}-e_g$ hopping. Magnetic susceptibility and spin wave 
spectra are calculated and compared to the experimental data, obtaining 
thereby the spin coupling constants in Na$_2$IrO$_3$ and 
Li$_2$IrO$_3$.
\end{abstract}

\date{\today}

\maketitle

In the quest for the materials with novel electronic phases, iridium oxide
Na$_2$IrO$_3$ came into focus
recently~\cite{Sin10,Tak10,Liu11,Cho12,Ye12,Sin12,Com12} due to theoretical
predictions~\cite{Jac09,Shi09} that this system may host Kitaev model physics
and quantum spin Hall effect. 

Na$_2$IrO$_3$ is an insulator with sizable and temperature independent optical
gap $\simeq 0.35$~eV~\cite{Com12}, and shows Curie-Weiss type
susceptibility~\cite{Sin10,Sin12} with moments corresponding to effective spin
one-half of Ir$^{4+}$ ion with $t_{2g}^5$ configuration~\cite{Abr70}. These
facts imply that Na$_2$IrO$_3$ is a Mott insulator with well localized
Ir-moments.

Collective behavior of local moments in Mott insulators is governed by three
distinct and often competing forces: ({\it i}) orbital-lattice (Jahn-Teller)
coupling, ({\it ii}) virtual hopping of electrons across the Mott gap
resulting in exchange interactions, and ({\it iii}) relativistic spin-orbit
coupling (see Ref.~\cite{Kha05} for extensive discussions). The corresponding
energy scales $E_{JT}, J$, and $\lambda$ vary broadly depending on the type of
magnetic ions and chemical bonding~\cite{Goo63}. When $\lambda>(E_{JT},J)$, as
often realized for Co, Rh, Ir ions in octahedral environment, local moments
acquire a large orbital component which may result in a strong departure from
spin-only Heisenberg models~\cite{Kha05,Jac09}. The direct observation of large
spin-orbit splitting $3\lambda/2\sim 0.6-0.7$~eV in insulating iridates
Sr$_2$IrO$_4$~\cite{JK214}, Sr$_3$Ir$_2$O$_7$~\cite{JK327}, and
Na$_2$IrO$_3$~\cite{YJK12} made it certain that $\lambda>(E_{JT},J)$. Thus,
low-energy physics of Na$_2$IrO$_3$ is governed by interactions among the
spin-orbit entangled Kramers doublets of Ir-ions. 

It is also established now~\cite{Liu11,Cho12,Ye12} that Ir-moments in
Na$_2$IrO$_3$ undergo antiferromagnetic (AF) order at $T_N\simeq 15$~K.
The fact that $T_N$ is much smaller than paramagnetic Curie temperature
($-125$~K)~\cite{Sin12} and spin-wave energies~\cite{Cho12} implies that the
underlying interactions are strongly frustrated. This is natural in so-called
Kitaev-Heisenberg (KH) model~\cite{Cha10} where long range order is suppressed
by the proximity to the Kitaev spin-liquid (SL) state. However, the observed
``zigzag'' magnetic pattern [ferromagnetic (FM) zigzag chains, AF-coupled to
each other] came as a surprising challenge to this simple and attractive
model. To resolve the ``zigzag puzzle'', various proposals, ranging from
routine extensions of the KH model~\cite{Kim11,Sin12,Bha12} to a complete
denial~\cite{Maz12} of the presence of local Ir-moments, have been put
forward.

In this Letter, we show that the zigzag order is in fact a natural ground
state (GS) of the KH model, in a previously overlooked parameter range. 
Next, we identify the exchange process that supports a zigzag-phase regime. 
Further, we calculate spin-wave spectra and magnetic susceptibility of the
model in zigzag phase, and find a nice agreement with 
experiments~\cite{Cho12,Sin10,Sin12}. This lends strong support to the 
KH model as a dominant interaction in Na$_2$IrO$_3$ and related oxides. 

{\it The model}.-- Nearest-neighbor (NN) interaction between isospin one-half 
Kramers doublets of Ir$^{4+}$ ions, coupled via 90$^\circ$-exchange bonds,
reads as follows (the exchange processes are described later): 
\begin{equation}
{\cal H}_{ij}^{(\gamma)}=2K~S_{i}^{\gamma}S_{j}^{\gamma}+
J~{\vc S}_i\!\cdot\!{\vc S}_{j}~.
\label{eq:KHmodel}
\end{equation}
Here, $\gamma(=x,y,z)$ labels 3 distinct types of NN bonds of a honeycomb 
lattice~\cite{Cha10} of Ir ions in Na$_2$IrO$_3$, and spin axes
are oriented along the Ir-O bonds of IrO$_6$ octahedron. The bond-dependent 
Ising coupling between the $\gamma$ components of spins is nothing but Kitaev 
model~\cite{Kit06}, and the second term stands for the Heisenberg exchange. 

Let us introduce the energy scale $A=\sqrt{K^2+J^2}$ and the angle $\varphi$ 
via $K=A\sin\varphi$ and $J=A\cos\varphi$; the model~\eqref{eq:KHmodel} takes
then the following form: 
\begin{equation}
{\cal H}_{ij}^{(\gamma)}=A\;(2\sin\varphi~S_{i}^{\gamma}S_{j}^{\gamma}+
\cos\varphi~{\vc S}_i\!\cdot\!{\vc S}_{j}).
\label{eq:model0}
\end{equation}
We let the ``phase'' angle $\varphi$ to vary from 0 to $2\pi$, uncovering
thereby additional phases of the model that escaped attention
previously~\cite{Cha10}, including its zigzag ordered state which is of a
particular interest here. 

It is highly instructive to introduce, following Refs.~\cite{Kha05,Cha10}, 4
sublattices with the fictitious spins $\tilde{\vc S}$, which are obtained from
${\vc S}$ by changing the sign of its two appropriate components depending on
the sublattice index. This transformation (generic for triangular, honeycomb,
kagome lattices) results in the $\tilde{\vc S}$-Hamiltonian of the same form
as \eqref{eq:KHmodel}, but with effective couplings $\tilde K=K+J$ and $\tilde
J=-J$, revealing a hidden $SU(2)$ symmetry of the model at $K=-J$ (where the
Kitaev term $\tilde K$ vanishes). For the angles, the mapping reads as
$\tan{\tilde\varphi}=-\tan\varphi-1$.        

\begin{figure}[tb]
\begin{center}
\includegraphics[width=8.0cm]{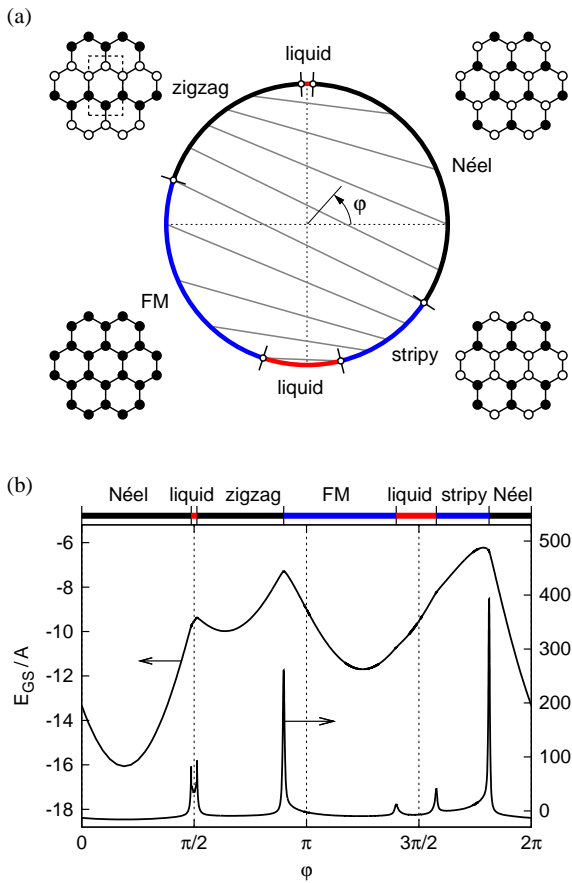}
\caption{(color online). 
(a) Phase diagram of the Kitaev-Heisenberg model containing 2 spin-liquid and 
4 spin-ordered phases. The transition points (open dots on $\varphi$-circle) 
are obtained by an exact diagonalization. The gray lines inside the circle 
connect the points related by the exact mapping (see text). Open/solid 
circles in the insets indicate up/down spins. The rectangular box in 
zigzag pattern (top-left) shows the magnetic unit cell. 
(b) Groundstate energy $E_\mathrm{GS}$ of 24-site cluster and its second 
derivative $-\mathrm{d}^2 E_\mathrm{GS}/\mathrm{d}\varphi^2$ revealing the phase
transitions.
}
\label{fig:phases1}
\end{center}
\end{figure}

{\it Phase diagram}.-- In its full parameter space, the KH model accommodates
6 different phases, best visualized using the phase-angle
$\varphi$ as in Fig.~\ref{fig:phases1}(a). In addition to the previously
discussed~\cite{Cha10,Jia11,Reu11} N\'eel-AF, stripy-AF, and SL states near
$\varphi=0$, $-\frac{\pi}{4}$, and $-\frac{\pi}{2}$, respectively, we
observe 3 more states. First one is ``AF'' ($K>0$) Kitaev spin-liquid near
$\varphi=\frac{\pi}{2}$. Second, FM phase broadly extending over
the third quadrant of the $\varphi$-circle. The FM and stripy-AF states are
connected [see Fig.~\ref{fig:phases1}(a)] by the 4-sublattice transformation,
which implies their identical dynamics. Finally, near
$\varphi=\frac{3}{4}\pi$, the most wanted phase, zigzag-AF, appears occupying
almost a quarter of the phase space. Thanks to the above mapping, it is
understood that the zigzag and N\'eel states are isomorphic, too. In
particular, the $\varphi=\frac{3}{4}\pi$ zigzag is identical to Heisenberg-AF
of the fictitious spins.  

To obtain the phase boundaries, we have diagonalized the model numerically,
using a hexagonal 24-site cluster with periodic boundary conditions. The
cluster is compatible with the above 4-sublattice transformation and $\varphi
\leftrightarrow \tilde\varphi$ mapping. As seen in Fig.~\ref{fig:phases1}(b),
the second derivative of the GS energy $E_\mathrm{GS}$ with respect to
$\varphi$ well detects the phase transitions.  Three pairs of linked
transition points are found: $(87.7^\circ, 92.2^\circ)$ and $(-76.1^\circ,
-108.2^\circ)$ for the spin liquid/order transitions around
$\pm\frac{\pi}{2}$, and $(161.7^\circ,-33.8^\circ)$ for the transitions
between ordered phases. 

The transitions from zigzag-AF to FM, and from stripy-AF to N\'eel-AF are of
first order by symmetry; see very sharp peaks in Fig.~\ref{fig:phases1}(b).
The spin liquid/order transitions near $\varphi=-\frac{\pi}{2}$ lead to wider
and much less pronounced peaks, suggesting a second (or weakly first) order
transition~\cite{Cha10}. On the contrary, liquid/order transitions around
$\varphi=\frac{\pi}{2}$ show up as very narrow peaks; on the finite cluster
studied, they correspond to real level crossings. Nature of these quantum
phase transitions remains to be clarified. 

While at $J=0$ (i.e. $\varphi=\pm\frac{\pi}{2}$) the sign of $K$ is
irrelevant~\cite{Kit06}, the stability of the AF- and FM-type Kitaev 
spin-liquids
against $J$-perturbation is very different: the SL phase near $\frac{\pi}{2}$
($-\frac{\pi}{2}$) is less (more) robust. This phase behavior is related to
a different nature of the competing ordered phases: for the $\frac{\pi}{2}$ SL,
these are highly quantum zigzag and N\'eel states, while the SL near
$-\frac{\pi}{2}$ is sandwiched by more classical (FM and ``fluctuation free''
stripy~\cite{Cha10}) states which are energetically less favorable than
quantum SL state.  
 
\begin{figure}[tb]
\begin{center}
\includegraphics[width=8.5cm]{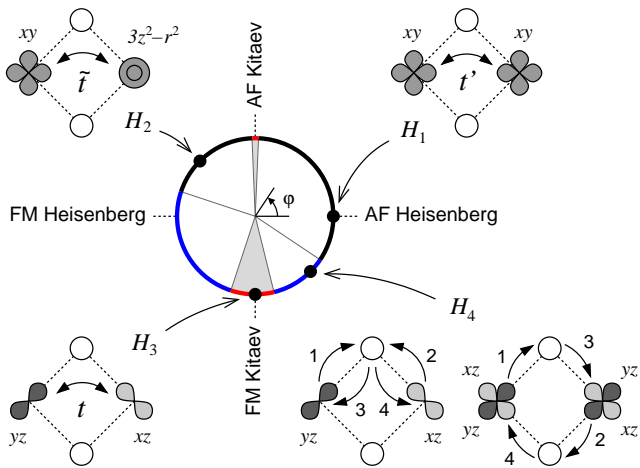}
\caption{(color online). 
Schematics of 4 different exchange processes (see text for details), arranged
around the $\varphi$-phase diagram of Fig.~\ref{fig:phases1}(a).  Taken 
separately, the Hamiltonians $H_1$, $H_2$, $H_3$, and $H_4$ would favor ``pure''
N\'eel-AF, zigzag-AF, Kitaev-SL, and stripy-AF states, respectively, as
indicated by arrows connecting $H_i$ with the dots on $\varphi$-circle. The
circle is divided into the phase-sectors by gray lines; SL phases are shaded.
}
\label{fig:interactions}
\end{center}
\end{figure}

{\it Exchange interactions in Na$_2$IrO$_3$}.-- Having fixed the parameter 
space $(K>0, J<0)$ for zigzag phase, we turn now to the physical processes 
behind the model~\eqref{eq:KHmodel}. Interactions between local moments in 
Mott insulators arise due to virtual hoppings of electrons. This may happen 
in many different ways, depending sensitively on chemical bonding, intra-ionic 
electron structure, etc. The case of present interest (i.e., strong spin-orbit 
coupling, $t_{2g}^5$ configuration, and $90^\circ$-bonding geometry) has been 
addressed in several papers~\cite{Kha05,Che08,Jac09,Cha10}. There are  
following four physical processes that contribute to $K$ and $J$ couplings. 

{\it Process 1}: Direct hopping $t^\prime$ between NN $t_{2g}$ orbitals. Since
no oxygen orbital is involved, $90^\circ$-bonding is irrelevant; the resulting
Hamiltonian is $H_1=I_1\;{\vc S}_i\!\cdot\!{\vc S}_{j}$ with
$I_1\simeq(\frac{2}{3}t^{\prime})^2/U$~\cite{Cha10}. Here, $U$ is Coulomb
repulsion between $t_{2g}$ electrons. Typically, one has $t^\prime/t<1$, when
compared to the indirect hopping $t$ of $t_{2g}$ orbitals via oxygen ions. 

{\it Process 2}: Interorbital NN $t_{2g}-e_g$ hopping $\tilde t$. This is 
the dominant pathway in $90^\circ$-bonding geometry since it involves strong 
$t_{pd\sigma}$ overlap between oxygen-$2p$ and $e_g$ orbital; typically, 
${\tilde t}/t\sim 2$. The corresponding Hamiltonian is~\cite{Kha05}: 
\begin{equation}
H_{2}^{(\gamma)}=I_2\;(2S_{i}^{\gamma}S_{j}^{\gamma}-{\vc S}_i\!\cdot\!{\vc S}_{j}). 
\label{eq:H2}
\end{equation}
This is nothing but the model~\eqref{eq:KHmodel} with $K=-J=I_2>0$, i.e., at
its SU(2) symmetric point $\varphi=\frac{3}{4}\pi$ inside the zigzag phase,
see Fig.~\ref{fig:interactions}. For the Mott-insulating iridates (as opposed
to charge-transfer cobaltates~\cite{Kha05}), we estimate $I_2\simeq
\frac{4}{9}(\tilde t/\tilde U)^2 \tilde J_H$, where $\tilde U$ is (optically
active) excitation energy associated with $t_{2g}-e_g$ hopping, and $\tilde
J_H$ is Hund's interaction between $t_{2g}$ and $e_g$ orbitals.  The physics
behind this expression is clear: $(\tilde t/\tilde U)^2$ measures the amount
of $t_{2g}$ spin which is transferred to NN $e_g$ orbital; once arrived, it
encounters the ``host'' $t_{2g}$ spin and has to obey the Hund's rule. 

For its remarkable properties, the Hamiltonian $H_2$~\eqref{eq:H2} deserves a
few more words. On a triangular lattice, it shows a nontrivial spin vortex
ground state (see Fig.~5 of Ref.~\cite{Kha05}); however, the elementary
excitations are simple $SU(2)$ magnons of a conventional Heisenberg-AF. When
regarded as ``$J$''-part of a doped $t-J$ model, it leads to an exotic
pairing~\cite{Kha04,Kha05}.     

{\it Process 3}: Indirect hopping $t$ between NN $t_{2g}$ orbitals via two
intermediate oxygen ions. This gives rise to the Kitaev model 
$H_{3}^{(\gamma)}=-I_3S_{i}^{\gamma}S_{j}^{\gamma}$, with 
$I_3\simeq\frac{8}{3}(t^2/U)(J_H/U)$~\cite{Jac09} where $J_H$ is Hund's 
coupling between $t_{2g}$ electrons. This process supports 
$\varphi=-\frac{\pi}{2}$ SL state, see Fig.~\ref{fig:interactions}. 
 
{\it Process 4}: Mechanisms involving $pd$ charge-transfer 
excitations (energy $\Delta_{pd}$). Two holes may meet at an oxygen 
(and experience $U_p$ repulsion), 
or cycle around a Ir$_2$O$_2$ plaquette (Fig.~\ref{fig:interactions}). 
The corresponding Hamiltonian $H_4$~\cite{Kha05,Che08,Jac09} has the form 
as of $H_2$~\eqref{eq:H2}. The coupling constant 
$I_4\simeq\frac{8}{9}(\frac{1}{\Delta_{pd}+U_p/2}-\frac{1}{\Delta_{pd}})$
turns out to be negative because of the near cancellation of the two 
terms~\cite{Jac09,noteI4}. It thus supports stripy-AF not observed 
in Na$_2$IrO$_3$. 
 
Putting things together, we observe that it is the interorbital $t_{2g}-e_g$ 
hopping $H_2$ process that uniquely supports zigzag order. This implies also 
that multiorbital Hubbard-type models, when applied to iridates with 
$90^\circ$-bonding geometry, must include $e_g$ states as well, even though 
the moments reside predominantly in the $t_{2g}$ shell. 

Up to this point, we neglected trigonal field splitting $\Delta$ of the 
$t_{2g}$ level due to the $c$-axis compression present in Na$_2$IrO$_3$. This 
approximation is valid as long as $\Delta$ is much smaller than spin-orbit 
coupling $\lambda\simeq 0.4$~eV~\cite{JK214,Sch84} and seems to be justified, 
since the recent {\it ab-initio} calculations~\cite{Maz12} suggest that 
$\Delta\simeq 75$~meV only~\cite{noteCos}.  

We have also examined the longer-range couplings, using the hopping matrix of 
Ref.~\cite{Maz12}, and found that second-NN interaction has the form 
of \eqref{eq:H2} (as noticed previously~\cite{Reu12}), while third-NN 
interaction is of AF-Heisenberg type. The second (third)-NN interaction would 
oppose (support) zigzag order; however, we believe that these couplings are not 
significant in Na$_2$IrO$_3$ because the corresponding long-range hoppings are
found to be small~\cite{Maz12,noteJ23}.   

We do not attempt here to evaluate the parameters involved in $H_1$--$H_4$; 
{\it ab-initio} calculations as in Ref.~\cite{Kat12} might be more useful in
this regard. Instead, having obtained a zigzag order in our 
model~\eqref{eq:KHmodel} and identified the physical process driving 
this order, we turn now to the experimental data. The $J$ and $K$ 
values in Na$_2$IrO$_3$ and Li$_2$IrO$_3$ will be extracted below from 
analysis of the neutron scattering and magnetic susceptibility data.  

\begin{figure}[tb]
\begin{center}
\includegraphics[width=8.5cm]{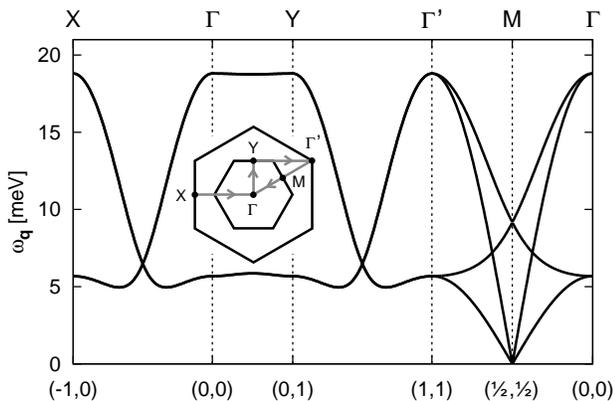}
\caption{
Magnon spectra in the zigzag phase calculated using
Eq.~\eqref{eq:spinwaves} with $(J,K)=(-4.01,10.45)$~meV.
The inset shows the path along the symmetry directions in the reciprocal
space; notations of Ref.~\cite{Cho12} are used.
}
\label{fig:spinwaves}
\end{center}
\end{figure}

{\it Spin-waves in the zigzag phase}.-- Consider a single domain zigzag state,
e.g., with FM chains running perpendicular to $z$-type bonds. Following
Ref.~\cite{Cho12}, we introduce a rectangular $a\times b$ magnetic unit cell
[$\sqrt{3}a_0\times 3a_0$ in terms of hexagon-edge $a_0$, see
Fig.~\ref{fig:phases1}(a)], and define the $ab$-plane wave vector $\bf q$ in
units of $(h,k)$ as ${\bf q}=(\frac{2\pi}{a} h, \frac{2\pi}{b} k)$. Standard
spin-wave theory gives four dispersive branches:  
\begin{eqnarray}
\omega_{1,2}^2(h,k)=\left[K^2+(K+J)^2\right]c^2_h - KJ(1-s_hs_k) \nonumber\\
\pm |(K+J)c_h|\sqrt{(2K-J)^2-(2Ks_h-Js_k)^2}\;,
\label{eq:spinwaves}
\end{eqnarray}
and $\omega_{3,4}(h,k)=\omega_{1,2}(-h,k)$, with $c_h=\cos\pi h$, 
$s_h=\sin\pi h$, and $s_k=\sin\pi k$. If $K=-J$, i.e. at 
$\varphi=\frac{3}{4}\pi$ point of hidden $SU(2)$ symmetry, two branches are 
degenerate ($\omega_{1}=\omega_{2}$) and become true Goldstone modes. Away 
from this special point, small magnon gap is expected to open by quantum 
effects not considered here. For $\bf q$ with $h=k$, the spin-wave 
dispersions~\eqref{eq:spinwaves} simplify to 
$\omega_{1}(h,h)=\sqrt{2K(2K+J)}\;|c_h|$ and $\omega_{2}(h,h)=\sqrt{2}|Jc_h|$,
revealing two different energy scales in magnon spectra set by $K$ and $J$
couplings. 

While the bandwidth of the lowest dispersive mode (set by $J$) is already known 
to be about 5-6 meV~\cite{Cho12}, we are not aware of the high energy magnon 
data to estimate $K$ in Na$_2$IrO$_3$. We have therefore examined (see below) 
the magnetic susceptibility data~\cite{Sin10,Sin12}, and obtained 
$(J,K)=(-4.01,10.45)$~meV that well fit the susceptibility as well 
as neutron scattering data~\cite{Cho12}. With this, we predict magnon 
spectra for Na$_2$IrO$_3$ shown in Fig.~\ref{fig:spinwaves}.  
The lowest dispersive ($J$) mode is as observed~\cite{Cho12}, indeed. 
However, mapping out entire magnon spectra is highly desirable to quantify 
the Kitaev term $K$ directly.     
 
\begin{figure}[tb]
\begin{center}
\includegraphics[width=8.5cm]{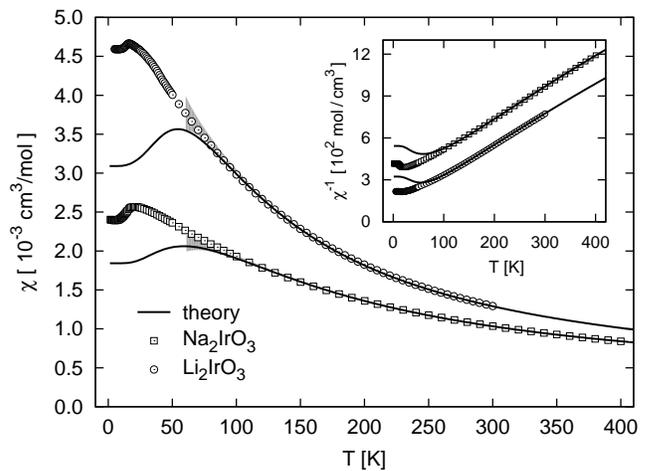}
\caption{
Experimental magnetic susceptibilities for Na$_2$IrO$_3$~\cite{Sin10,Sin12}
(squares) and Li$_2$IrO$_3$~\cite{Sin12} (circles) fitted by theoretical
results calculated with $(J,K)=(-4.01,10.45)$~meV and $(-5.30,7.85)$~meV,
respectively. Exact $\chi$ of the 8-site cluster is shown as solid lines.
Lanczos results for the 24-site cluster are indicated by
shading~\cite{noteFTLM}. Their comparison suggests that the calculated $\chi$
gives the thermodynamic limit down to $T\approx 100\:\mathrm{K}$ where the
finite-size effects become significant.
}
\label{fig:susceptibility}
\end{center}
\end{figure}

{\it Magnetic susceptibility }.-- We have calculated the uniform magnetic
susceptibility $\chi(T)$ of the model~\eqref{eq:KHmodel} on 8-site cluster
(using full exact diagonalization) and 24-site cluster (using
finite-temperature Lanczos method~\cite{Jak00,noteFTLM}). Both clusters are
compatible with the zigzag order when periodic boundary conditions are
applied. The parameters are varied such that $J=A\cos\varphi$ is consistent
with the neutron data~\cite{Cho12} while $\varphi$ stays within the zigzag
sector of Fig.~\ref{fig:phases1}(a); this strongly narrows the possible
$K$-window. For the data fits, we let $g$-factor of Ir$^{4+}$ ion to deviate
from 2 (due to the covalency effects~\cite{Abr70}), and include
$T$-independent Van Vleck term $\chi_0$. The result for $J=-4.01$ meV,
$K=10.45$ meV, $g=1.78$, $\chi_0=0.16\times 10^{-3}$cm$^3$/mol fits the
Na$_2$IrO$_3$ data nicely (Fig.~\ref{fig:susceptibility}); deviations occur at
low temperatures only, when correlation length exceeds the size of the cluster
used. The fit is quite robust: similar results can be found for small only
variations, locating Na$_2$IrO$_3$ near $\varphi=111 \pm 2^\circ$ of the model
phase diagram Fig.\ref{fig:phases1}(a). The spin couplings obtained are
reasonable for the $90^{\circ}$-exchange bonds (as
expected~\cite{Jac09,Kha05}, they are much smaller than in $180^{\circ}$-bond
perovskites~\cite{JK214,JK327}).  The magnitude of Van Vleck term also agrees
with our estimate $\chi_0\simeq\frac{8}{3\lambda}\mu_{\rm B}^2N_{\rm A}\simeq
0.2\times 10^{-3}$ cm$^3$/mol for Ir$^{4+}$ ion, considering spin-orbit
coupling $\lambda\simeq 0.4$~eV~\cite{JK214,Sch84}.

For the sake of curiosity, we have also fitted $\chi(T)$ data of
Li$_2$IrO$_3$~\cite{Sin12}, a sister compound  of Na$_2$IrO$_3$. Acceptable
results have been found for the angle window $\varphi=124 \pm 6^\circ$; a
representative plot for $J=-5.30$ meV, $K=7.85$ meV, $g=1.94$,
$\chi_0=0.14\times 10^{-3}$ cm$^3$/mol is shown in
Fig.~\ref{fig:susceptibility}.  It is worth noticing that the value of $J$,
which controls the bandwidth of the softest spin-wave mode (see
Fig.~\ref{fig:spinwaves}), appears to be similar in both compounds. This may
explain why they undergo magnetic transition at similar $T_N\simeq15$~K,
despite very different high temperature susceptibilities. 

To conclude, we have clarified the origin of zigzag magnetic order in 
Na$_2$IrO$_3$ in terms of nearest-neighbor Kitaev-Heisenberg model for
localized Ir-moments. The model well agrees with the low-energy magnon 
and high temperature magnetic susceptibility data. A general implication 
of this work is that the interactions considered here should hold a
key for understanding the magnetism of a broad class of spin-orbit Mott
insulators with $90^{\circ}$-exchange bonding geometry, including triangular, 
honeycomb, hyperkagome lattice iridates. 

We thank R. Coldea, Y. Singh, and H. Takagi for discussions.
JC acknowledges support by the Alexander von Humbolt Foundation,
ERDF under project CEITEC (CZ.1.05/1.1.00/02.0068), and EC 7$^{\rm{th}}$
Framework Programme (286154/SYLICA). GJ is supported by GNSF/ST09-447 and
in part by the NSF under Grant No. NSF PHY11-25915.


\begin{thebibliography}{99}
\bibitem{noteJ23} E.g., the second-NN hopping $\sim75$~meV is about 4 times 
less than NN hopping $t$ via oxygen~\cite{Maz12}; note that hopping matrix 
elements enter in spin couplings quadratically.
\bibitem{noteFTLM}
The finite-temperature Lanczos method was performed using $M=100$ Lanczos
steps and $N_\mathrm{st}=1024$ random sampling vectors. The values of $\chi$
and their statistical error are presented in Fig.~\ref{fig:susceptibility}
in the form of $3\sigma$-intervals estimated by taking many different sets 
of the sampling vectors.
\end{thebibliography}
\end{document}